\documentclass[pra,twocolumn,showpacs,preprintnumbers,amsmath,amssymb]{revtex4}
\usepackage{mathrsfs}
\usepackage{bbm}
\usepackage{amssymb}
\usepackage{amsfonts}
\usepackage{tipa}
\usepackage[colorlinks,linkcolor=blue,hyperindex,CJKbookmarks]{hyperref}
\usepackage{epsfig,graphicx}
\usepackage{amstext}
\usepackage{amsmath}

\newcommand{\e}{\textmd{e} }
\newcommand{\tr}[1]{{\rm{tr}}\left[#1\right]} 
\begin{document}
\title{Improved Landauer's principle and generalized second law of thermodynamics with initial correlations and non-equilibrium surrounding environments}

\author{Ke-Xia Jiang}
\email{kexiajiang@126.com}
\author{Yuan-Mou Li}
\affiliation{Department of Physics, Engineering University of CAPF, Xi'an 710086, P. R. China}

\author{Heng Fan}
\email{hfan@iphy.ac.cn}
\affiliation{Beijing National Laboratory for Condensed Matter Physics,
               Institute of Physics, Chinese Academy of Sciences, Beijing
               100190, P. R. China}
\date{\today}
\begin{abstract}
Traditional form of the second law of thermodynamics is strongly restricted by three conditions: One is the initial joint state of the system and surroundings should be a product state, so that there exists no initial correlations. The second is the initial states of surroundings are in equilibrium thermodynamics. And the end is weak couplings between the system and surroundings. This formulation of the second law should be reexamined in order to understand the relations of thermodynamics and information theory, especially, when existing initial correlations. In this work, using the techniques of quantum statistical mechanics for thermodynamics and quantum information science, we recast fundamental laws of thermodynamics from theoretical information point of view. Initial correlations between the system and surroundings are considered, which evolves thermodynamically and result in modifications of the traditional formulations. We obtained improved forms of the entropy increase, Landauer's principle, and the second law of thermodynamics, which are exhibited as equalities rather than inequalities, and through which physical nature of information can be demonstrated precisely. Further, using the language totally belongs to quantum information theory, we give the direction of natural evolution process a new statement: the evolution of an isolated quantum system, where no correlations exist between subsystems, initially, always towards to directions of the correlation information never be decreased. Such result indicates that the traditional principle of entropy increase can be redescribed using information theory, identically.
\end{abstract}
\maketitle
\section{Introduction}\label{introduction}
The Maxwell's demon paradox \cite{Maxwell1871} leads a long history of controversy for the meaning of information in physics, especially in thermodynamics \cite{MaxwellsDemon}. After all, the formulation of the second law of thermodynamics,
attributed to Rudolf Clausius, Lord Kelvin and Max Planck, makes no mention of information \cite{Parrondo2015}. An outstanding insight came from Rolf Landauer in 1961 \cite{Landauer1961}. The definitive discovery, which now called Landauer's principle, asserts the relationship between information and physics: the erasure of information on the state of a system is concomitant with the dissipation of heat into the surroundings. Further, thanks to Bennett's efforts \cite{Bennett1982} in clarifying the role of Maxwell's demon
in the second law of thermodynamics which was disturbed deeply.
Such physical nature of information inspires great interest to explore their inherent interconnections. Obviously,
it is a fundamental task to exploit relationships between information theory and thermodynamics for physics. Solving the following two tasks is especially necessary \cite{Parrondo2015}: refining the second law of thermodynamics to incorporate information explicitly, and clarifying the physical nature of information, so that it enters the second law not as an abstraction, but as a physical entity.

Since Maxwell's demon, great progress have been made in investigating extensively the relationships between information theory and thermodynamics both theoretically and experimentally (see~\cite{Maruyama2009,Parrondo2015,Gooldetal2015,Brandao2015,Koski2015,Lutz2015,Berut2015,Millen2016,Vinjanampathy2016,Sagawa2017} for a recent overview).
A remarkable development is in the field of non-equilibrium statistical mechanics, where fluctuation theorems~\cite{Harris2007,Jarzynski2011,Esposito2009,Goold2015} and stochastic thermodynamics clarified and generalized the classical results~\cite{Campisi2011,Seifert2012}. And more importantly, they present promising routes to understand the relationship between information and thermodynamics in general frameworks. Through various perspectives and researches, close relationships between quantum information and thermodynamics show in many other aspects, such that, work extraction problems from quantum systems \cite{Horodecki2013,Skrzypczyk2014,Frenzel2014,Faist2015,Binder2015}, and thermodynamics meanings of quantum correlations or coherence \cite{Rio2011,Jevtic2012,Lostaglio2015,Kammerlander2016}, further, the resource theory of quantum information in thermodynamics \cite{Gour2015}.
 Recently, different from the previous methods, Reeb and Wolf  \cite{Reeb2014} using techniques of quantum statistical mechanics, provided a rigorous proof of an improved version of Landauer's principle from the underlying microscopic equations, which is formulated in terms of an equality rather than inequality, impressively.

Sagawa and Ueda considered the role of initial correlations in stochastic thermodynamics when establishing a more generalized fluctuation theories~\cite{SagawaUeda2012}. However, most work in literatures in investigating relations between information and thermodynamics assume that subsystems are independent and the initial correlations between them are ignored.

And very recently, excellent works~\cite{Strasberg2017,Bera2017} are attempted to extend the traditional framework of thermodynamics, especially when correlations are exist initially. However, a clarified and well-accepted expression about the second law of thermodynamics which incorporates quantum information theory is still challenging. And further, a natural and crucial question is: whether the traditional forms of thermodynamics, such as the principle of entropy increase and the second law, have another meanings belongs totally to (quantum) information theory, since they are so closely connected? By virtue of such expression, we may clarify their correlations clearly.

In this paper, using the techniques of quantum statistical mechanics put forward by Reeb and Wolf~\cite{Reeb2014}, we recast relations between the information theory and the fundamental laws of thermodynamics in a more general way, and establish a generalized frame to connect the two theories. Main developments are concentrated in three aspects: Firstly, the role of initial correlations between the system and surroundings are being considered. Secondly, from three versions which seem totally different, we obtain their improved formulas, where correlations play key roles. Actually, all the three versions we discussed can be viewed as descriptions of the second law of thermodynamics in different setups. We obtain more general forms of the second law of quantum thermodynamics which incorporate information explicitly and naturally. Finally, we discuss the direction of natural evolution process. We find it can be described by using the language totally belongs to quantum information theory.

This paper is organized as follows: in Sec.~\ref{sec.Infandthermo}, we first discuss the concepts of information entropy and thermodynamic entropy and give out a fundamental Lemma which is at the heart of the whole discussions. Taking into account the role of initial correlations between the system and the bath reservoir, where the state of the surrounding environments maybe far away from equilibrium, in Sec.~\ref{sec.evolution}, we obtain our main results: the improved theories for quantum evolution dynamics. In Sec.~\ref{sec.examples}, we illustrate the results through a specific physical example in open-system dynamics. The final section, Sec.~\ref{conclusions}, is devoted to conclusions and discussions.

\section{Two concepts of entropy}\label{sec.Infandthermo}
Two concepts of entropy in the quantum field are fundamentally important in the whole proceedings, von Neumann entropy $S_V(\rho)$ (quantum information entropy) and thermodynamic entropy $S_{\rm{th}}(\rho)$. Von Neumann entropy has the meaning belongs to category of quantum statistical mechanics from its originally assuming, as an extension of classical Gibbs entropy concept to the field of quantum mechanics, which is introduced by John von Neumann around 1927 for proving the irreversibility of quantum measurement processes \cite{Neumann1932}.
In quantum information theory, von Neumann entropy
\begin{equation}\label{vonNeumannentropy}
 S_V(\rho)=-\tr{ \rho \log \rho},
\end{equation}
namely, quantum information entropy, appears as the analogue of classical Shannon entropy \cite{Wehrl1978,Vedral2002,Cover2006}. The concept is critical important for more general questions in the affinity of the two theories, quantum information theory and thermodynamics \cite{Henderson2003}.

In literatures, controversies regarding the relationship between the two concepts exist~\cite{Jaynes1957,Brillouin1962,Zurek19891996,Berry2003,Henderson2003}. In specific fields of thermodynamics, which kind thermodynamic entropy is identified with information entropy should be carefully considered \cite{BerrySanders2003}. And the fact is that both of the two concepts stem from the similar statistical characteristics \cite{Wehrl1978}.

However, in the following analysis, we will use the viewpoint in Refs.~\cite{Brillouin1962,Parrondo2015} presenting Boltzmann's constant as a scale factor to distinguish the two concepts. Before setup, in order to highlight its basic and fundamental position, we propose the following Lemma.

 {\it Lemma 1}\label{Lemma1}: In quantum thermodynamic of information theory, thermodynamic entropy $ S_{\rm{th}}(\rho)$ of a physical state $\rho$  connects with von Neumann entropy  $S(\rho):=S_{V}(\rho)=-\tr{ \rho \log \rho}$ of the same state in which information memory are stored, as
 \begin{equation}\label{thermodynamicNeumann}
 S_{\rm{th}}(\rho)=k S_V(\rho),
 \end{equation}
with the Boltzmann's constant $k$ to balance the dimensions.

Although, the Lemma is not a new discovery, it states a noticeable characteristic: $S_{\rm{th}}(\rho)$ belongs to thermodynamics, while $S_{V}(\rho)$ belongs to information theory. Such a relation is fundamental and at the heart of the whole discussions. Loosely speaking, if do not care about the Boltzmann constant $k$ or put it equal to 1, the two concepts can be regarded as equivalent. In most literatures, this is a common convention and self-evident.

\section{Improved theories for quantum evolution dynamics}\label{sec.evolution}
We consider an isolated quantum system ${\cal S}$,  which has no information and no energy exchanges with the outside. Firstly, we suppose the system composed by subsystems ${{\cal S}_{\mu}}$, $\mu=1,2,\ldots,N$.

\subsection{Improved principle of entropy increase} \label{entropyincreaseprinciple}
Two different concepts of entropy, the joint entropy and the marginal entropy, which be defined globally and locally, respectively, should be clarified when considering the entropy of isolated quantum systems.

{\it  Definition 1}\label{ Definition1}: For a quantum system ${\cal S}$ of state $\rho$, the joint entropy is the von Neumann entropy of the total system, $S_{JE}(\rho)=S_V(\rho)$.

{\it  Definition 2}\label{ Definition2}: For a quantum system ${\cal S}$ of state $\rho$, which is composed by subsystems ${\cal S}_{\mu},  \mu=1,2,\ldots,N$, the sum of all von Neumann entropies of the reduced states $S_{ME}(\rho)=\sum_{\mu}{S_V(\rho_{\mu})}$ is called the marginal entropy of the system ${\cal S}$, with $\rho_{\mu}=\rm{tr}_{\hat{\mu}}{\left[\rho\right]}$ by tracing all subsystems except ${\cal S}_{\mu}$.

The above two concepts of entropy are in general different, except for the case $\rho$ is a product state for all subsystems, $\rho=\rho_{1}\otimes\rho_{2}\otimes\ldots\otimes\rho_{N}$. However, differences are sometimes being ignored
in statements of the second law of thermodynamics.

Another basic concepts in information theory is mutual information $I({\cal S:B})$~\cite{Cover2006} for two random variables, which essentially measures the amount of information that one variable contains about another. Mutual information is always non-negative.
In the following, our set up is based on a quantum system with Hilbert space $H$. Therefore, mutual information is the quantum mutual information $I({\cal S}:B)=I(\rho_{{\cal S}B})$, which is an essential measurement for evaluating total correlations between quantum subsystems~\cite{Vedral2002,Groisman2005,Schumacher2006,LiLuo2007,NielsenChuang2000}.
Correlation information \cite{Lindblad1973,Horodecki1994,Herbut2004} contained in a quantum system is a generalized concept of mutual information for  multi-random variables in quantum information theory. Using the above two definitions, correlation information, now can be expressed as,
\begin{equation}\label{correlationinf}
I{(\rho)}=S_{ME}(\rho)-S_{JE}(\rho),
\end{equation}
which is also nonnegative. Iff no correlations exist between all subsystems, it becomes zero, i.e. $I{(\rho)}=0$.

For an isolated quantum system, the joint entropy is always conserved, $S_{JE}(U^{\dag}\rho U)=S_{JE}(\rho)$, since the evolution can be described by unitary transformations $U=\e ^{-iHt}$. Straightforwardly, we have the following result.

{\it Theorem 1}\label{Theorem1}: During the evolution of an isolated quantum system ${\cal S}$ composed by subsystems ${\cal S}_{\mu}, \mu=1,2,\ldots,N$, from the initial state $\rho _{i}$ to the final state $\rho _{f}$, changes of thermodynamically marginal entropy are equal to changes of correlation information times Boltzmann's constant $k$,
\begin{equation}\label{Improvedentropyincrease}
\Delta S_{{\rm{th}},ME}:=k\Delta S_{ME}=k\Delta I=k \left [I(\rho_{f})- I(\rho_{i})\right ].
\end{equation}

Using  Lemma 1 and Eq.~\eqref{correlationinf}, the proof of Theorem 1 is straightforward and omitted. Theorem 1 both improves and generalizes the traditional explanation about entropy increase, we call it {\it the improved principle of entropy increase}.
The traditional form can be easily obtained by attaching additional conditions: if all subsystems ${\cal S}_{\mu}, \mu=1,2,\ldots,N$, are initially uncorrelated, i.e. $I(\rho_{i})=0$ iff $\rho_{i}=\rho_{1}\otimes\rho_{2}\otimes\ldots\otimes\rho_{N}$. We call it the ``product condition''. Then, thermodynamically marginal entropy of the system ${\cal S}$ will never be decreased,
 \begin{equation}\label{principleentropyincrease}
 \Delta S_{{\rm{th}},ME}=S_{{\rm{th}},ME}(\rho_f)-S_{{\rm{th}},ME}(\rho_i)\geq 0.
 \end{equation}

However, when initial correlations exist (see the illustration of Fig.~\ref{initialmutualinformation} in Sec.~\ref{sec.examples}), i.e. $I(\rho_{i})>0$, things may be different. Impressively, marginal entropy of the whole system can be decreased, $\Delta S_{{\rm{th}},ME}\leq 0$, due to the existence of initial correlations between the subsystems, where thermodynamic features of the negative information entropy are obviously expressed \cite{Rio2011}. From the illustrations (Fig.~\ref{changeratetotalentropy} in Sec.~\ref{sec.examples}), we can see that the change rate of total entropy can be negative in process of dynamical evolutions. Our results about relationships between the total entropy of the whole system and the mutual information are consistent with other discussions in references \cite{Lloyd1989,Groisman2005,SagawaUeda2012,SagawaUeda2013}, which used different methods.

We remark that it may be an interesting question to consider the initial correlations for other systems, such as universe. And a remarkable result is straightforward: If correlation information of the state of the total and isolated universe is constant, $I(\rho_U)\equiv {\rm{Const}}.$, then thermodynamically marginal entropy of the total universe is also constant, $S_{{\rm{th}},ME}\equiv {\rm{Const}}$.

\subsection{Improved Landauer's principle} \label{improvedLandauersprinciple}
The discussion in the above subsection is general and does not assume any particulary conditions for quantum processes. Now, we compact the total system by only two subsystems, ${\cal S}$ and ${\cal B}$, which denote the register system of information and a finite-sized environment or bath reservoir, respectively, and refine the setup.

Without loss generality, we consider an information erasing process, in which the (sub)system ${\cal S}$ with a time-dependent Hamiltonian $H_{\cal S}(t)$, in contact with the bath reservoir ${\cal B}$ with a time-independent Hamiltonian $H_{\cal B}$. The total Hamiltonian is
\begin{equation}\label{totalHamiltonian}
H(t)=H_{\cal S}(t) +H_{\cal B}+H_{\rm{int}}(t),
\end{equation}
where the last term is the coupling strength between ${\cal S}$ and ${\cal B}$. The time evolution of the initial state $\rho_{{\cal SB}}(0)$
can be written as
\begin{equation}\label{eqevolution}
\rho_{{\cal SB}}(t)=U(t)\rho_{{\cal SB}}(0)U(t)^\dagger.
\end{equation}
In the following discussions, we relabel the finial reduced states as $\rho^{\prime}_{\mu}:= \rho_{\mu}(t)= \rm{tr}_{ \nu} [ \rho_{\mu \nu}(t)]:=\rm{tr}_{ \nu} [ \rho^{\prime}_{\mu \nu}]$, conveniently, distinguishing from the initial reduced states  $\rho_{\mu}:= \rho_{\mu}(0)=\rm{tr}_{ \nu} [ \rho(0)_{\mu \nu}]:=\rm{tr}_{ \nu} [ \rho_{\mu \nu}]$, with $\mu,\nu={\cal S},{\cal B}$.

Limiting the bath reservoir always in an equilibrium state, sometimes is unnecessary, such as for small sized surrounding environments. One method to circumvent the obstacle is introducing a corresponding thermal state $\rho_{{\cal B},\rm{th}}=e^{-\beta H_{\cal B}}/Z$ for non equilibrium surroundings, with the partition function $Z=\tr{e^{-\beta H_{\cal B}}}$ and the inverse temperature $\beta=1/kT$, by assuming they have the same energy
\begin{equation}\label{sameenergy}
E_{\cal B}=\tr {\rho_{{\cal B},\rm{th}} H_{\cal B}}=\tr{\rho_{\cal B} H_{\cal B}}.
\end{equation}

According to Theorem 3 in Ref.~\cite{Reeb2014}, we can have a similar analysis
\begin{widetext}
\begin{align}\label{ap001}
 S(\rho^{\prime}_{\cal B})-S(\rho_{{\cal B},\rm{th}})=& S(\rho^{\prime}_{\cal B})+\tr{\rho_{{\cal B},\rm{th}} \log \frac{\e^{-\beta H_{\cal B}}}{\tr{\e^{-\beta H_{\cal B}}}}}\nonumber\\
 =& S(\rho^{\prime}_{\cal B})+\tr{\rho_{{\cal B},\rm{th}} \left( -\beta H_{\cal B} -\log \tr{\e^{-\beta H_{\cal B}}}  \right) }\nonumber\\
 =&S(\rho^{\prime}_{\cal B})-\beta \tr{\rho_{{\cal B},\rm{th}}H_{\cal B}}-\log \tr{\e^{-\beta H_{\cal B}}}+\beta \tr{\rho^{\prime}_{\cal B}H_{\cal B}}-\beta \tr{\rho^{\prime}_{\cal B}H_{\cal B}}\nonumber\\
 =&\beta \tr{H_{\cal B}\left(\rho^{\prime}_{\cal B}-\rho{_{\cal B}}\right)}+S(\rho^{\prime}_{\cal B})+\tr{\rho^{\prime}_{\cal B}\log \frac{\e^{-\beta H_{\cal B}}}{\tr{\e^{-\beta H_{\cal B}}}}}\nonumber\\
 =&\beta \Delta Q- D(\rho^{\prime}_{\cal B}||\rho_{{\cal B},\rm{th}}),
\end{align}
\end{widetext}
Here, $\Delta Q=\tr {H_{\cal B} (\rho^{\prime}_{\cal B}-\rho_{\cal B})}$ is the heat dissipated from the system ${\cal S}$ to the surrounding bath reservoir ${\cal B}$, and $\Delta S=S(\rho_{\cal S})-S(\rho^{\prime}_{\cal S})$ is the decrease of the information entropy of the system. Without loss generality, we assume the evolution is an erasing information process inside the subsystem ${\cal S}$, so that $\Delta S$ is positive. The relative entropy between two states $\sigma$ and $\rho$ is defined as $D(\sigma || \rho)=\tr{\sigma \log \sigma}-\tr{\sigma \log \rho}$.

When replacing $\rho^{\prime}_{\cal B}$ with $\rho_{\cal B}$, $\Delta Q$ becomes zero automatically. So we have
\begin{equation}\label{app003}
    S(\rho_{{\cal B},\rm{th}})-S(\rho_{\cal B})=D(\rho_{\cal B}||\rho_{{\cal B},\rm{th}}).
\end{equation}
The non equilibrium surroundings bath reservoir brings about a difference, which shows as relative entropy, is understandable. Using Eqs.~\eqref{Improvedentropyincrease}, \eqref{ap001} and \eqref{app003}, we obtain
\begin{align}\label{app001}
    \Delta S+ \Delta I &= S(\rho^{\prime}_{\cal B})- S(\rho_{\cal B})\nonumber\\
     &= S(\rho^{\prime}_{\cal B})-S(\rho_{{\cal B},\rm{th}})+S(\rho_{{\cal B},\rm{th}})-S(\rho_{\cal B})\nonumber\\
     &=\beta \Delta Q- D(\rho^{\prime}_{\cal B}||\rho_{{\cal B},\rm{th}})+D(\rho_{\cal B}||\rho_{{\cal B},\rm{th}}).
\end{align}
Denoting the increasing of the relative (information) entropy of the bath reservoir with its initial thermal state as
\begin{equation}\label{increasingrelen}
\Delta D= D(\rho^{\prime}_{\cal B}||\rho_{{\cal B},\rm{th}})-D(\rho_{\cal B}||\rho_{{\cal B},\rm{th}}).
\end{equation}
Despite the fact that the relative entropy is always non-negative, $\Delta D$ does not have such property due to non-equilibrium characteristics of the surroundings (see Fig.~\ref{changeraterelativeentropy} in Sec.~\ref{sec.examples} for a example of open-system dynamics evolutions).

Further, using Theorem 1 and Eqs.~\eqref{app001}, we can obtain a more general form of  the Landauer's principle,
\begin{equation}\label{improvedLandauerprinciple}
\frac{ \Delta Q}{T}=k\left[\Delta S+\Delta I+ \Delta D\right].
\end{equation}
This is a further improved formula incorporates and emphasizes the initial correlations and non-equilibrium characteristics of the bath reservoir. Consistent with and as a continuation of the result in Ref.~\cite{Reeb2014}, we call it {\it the improved Landauer's principle}.

The traditional form of the Landauer's principle can be easily deduced by imposing two additional initial conditions: One is the ``product condition''-the system and surrounding environment are uncorrelated, i.e. $\rho_{{\cal S}B}=\rho_{\cal S}\otimes\rho_{\cal B}$, which ensures $\Delta I\geq 0$. And the other we call the ``thermal equilibrium condition"-state of surrounding environment is an equilibrium thermal state, i.e. $\rho_{\cal B}=\rho_{{\cal B},\rm{th}}$, which ensures $\Delta D\geq 0$.  So, we have the traditional form of the Landauer's principle,
\begin{align}\label{traditionalLandauer}
\beta \Delta Q \geq \Delta S,
\end{align}
which states that the erasure of information represented as $\Delta S$ will cost at least heat dissipation $\Delta Q$.

Further, we can have an equality which quantized the fluxes of heat, entropy and the change rate of information
\begin{equation}\label{flowofheat}
\beta  \dot{Q}=\dot{S}+\dot{I}+ \dot{D},
\end{equation}
with the time change rate  $(\dot{\hspace*{1mm}}):=\rm{d}/\rm{d}t$. Such a general form clarifies contributions of fluxes of each part.
When imposing the product and thermal equilibrium conditions, we can obtain expression of the traditional form of Landauer's principle using heat and entropy fluxes, $\beta  \dot{Q}\geq\dot{S}$~\cite{additions}, discussed in~\cite{GrangerKantz2013,Lorenzo2015} based on the other models or perspectives.

On the other hand, strong initial correlations between the system and the bath may lead to negative values of  $ \Delta{I}< 0$, and further, violate the traditional form of Landauer's principle. Such abnormal behaviors would entail the establishment of memory effects within the environment and the occurrence of non-Markovianity in the ${{\cal S-B}}$ dynamics \cite{McCloskey2014,Bylicka2016,Lorenzo2015}.

In Fig.~\ref{changerate} (see Sec.~\ref{sec.examples}), we plot change rates of the mutual information plus the relative entropy of the bath reservoir, namely, $ \dot{I}+\dot{D}$. The illustration shows the role of the fluxes of surrounding environment and especially the mutual information, in information processing for the open-system dynamics by using the Landauer's principle. The traditional flux inequality of the Landauer's principle, $\beta  \dot{Q}\geq\dot{S}$, can be violated during evolution processes due to the negative value of  $ \dot{I}+\dot{D}$.

\subsection{Improved second law of thermodynamics} \label{generalizedsecondlaw}
The second law of thermodynamics in the original formulation of macroscopic thermodynamics stipulates that the work $W$, required to change the state of a system in contact with a bath reservoir between two different equilibrium states, is at least equal to the corresponding increase of free energy for states in equilibrium, namely,
\begin{equation}\label{secondlaw}
W-\Delta F\geq 0.
\end{equation}
However, for a general quantum information erasing (or writing) process, the case maybe totally different, for example, during which the bath reservoir states in non-equilibrium and importantly, and correlations between the system and bath reservoir exist initially.

According to the first law of thermodynamics, that is the energy conservation law, distributions of the amount of work $W$ done on the total system can be generally written as,
\begin{align}\label{distributionwork}
W=&\tr{H^{\prime}\rho^{\prime}_{\cal{{\cal SB}}}}-\tr{H\rho_{\cal{{\cal SB}}}} \nonumber\\
=&\tr{H^{\prime}_{\cal S}\rho^{\prime}_{{\cal S}}}- \tr{H_{\cal S}\rho_{{\cal S}}}+\tr{H^{\prime}_{\cal B}\rho^{\prime}_{\cal B}}- \tr{H_{\cal B}\rho_{\cal B}}\nonumber\\
&+\tr{H^{\prime}_{\rm{int}}\rho^{\prime}_{{\cal SB}}}- \tr{H_{\rm{int}}\rho_{{\cal SB}}}\nonumber\\
=&\Delta E_{{\cal S}}+\Delta E_{\cal B}+\Delta \mathcal{E}_{{\cal SB}}.
\end{align}
where $\Delta E_{\cal S}={\rm{tr}}_{{\cal S}}\left[\rho_{{\cal S}}^{\prime} H_{\cal S}^{\prime} -\rho_{{\cal S}} H_{\cal S} \right ]$ and since  in Eq.~\eqref{totalHamiltonian} $H_{\cal B}$ is time independent, $\Delta Q=\tr{H_{\cal B}(\rho_{\cal B}^{\prime}-\rho_{\cal B})}$ is the heat transferred from bath reservoir, and $\Delta \mathcal{E}_{{\cal SB}}=\tr {H_{{\rm int}}^{\prime}\rho^{\prime}-H_{{\rm int}}\rho}$ can be understood as the corresponding energy change for interacations between the two subsystems. In order to get desired results, we have assumed the work $W$ done only on the subsystem ${\cal S}$.

Finding a proper definition of free energy $F$ for general statistical states $\rho$,  has been discussed by many authors from different perspectives and applications \cite{Esposito2011,SagawaUeda2012,Reeb2014,Brandao2015}. Here we adopt the definition in~\cite{Parrondo2015,Esposito2010,Esposito2011,Crooks2007}
\begin{equation}\label{freeenergy}
F_H(\rho)=\tr{\rho H}-\beta^{-1} S(\rho),
\end{equation}
where $\beta=1/k T$ is the inverse temperature of the thermal bath reservoir. We can also relax the limitation of thermal equilibrium condition of the surroundings, although non equilibrium states may not have well-defined temperatures. Similar with the above subsection, we circumvent the obstacle using Eq.~\eqref{sameenergy}.

By virtue of Theorem 1, using Eqs.~\eqref{app001}, \eqref{distributionwork} and \eqref{freeenergy}, we have
\begin{widetext}
\begin{align}\label{deducesecondlaw}
   W- \beta^{-1}\Delta I =&\tr{\rho^{\prime}_{\cal S} H^{\prime}_{\cal S}}-\tr{\rho_{\cal S} H_{\cal S}}+\Delta Q+\Delta \mathcal{E}_{{\cal SB}} +\beta^{-1} \Delta S-\beta^{-1}\left[S(\rho^{\prime}_{\cal B})-S(\rho_{\cal B}) \right]\nonumber\\
   =&\left(\tr{\rho^{\prime}_{\cal S} H^{\prime}_{\cal S}}- \beta^{-1} S(\rho^{\prime}_{\cal S})\right)-\left(\tr{\rho_{\cal S} H_{\cal S}}-\beta^{-1} S(\rho_{\cal S})\right)+\Delta \mathcal{E}_{{\cal SB}}+\Delta Q-\beta^{-1}\left[S(\rho^{\prime}_{\cal B})-S(\rho_{\cal B}) \right] \nonumber\\
   =&F_{H^{\prime}_{\cal S}}(\rho^{\prime}_{\cal S})-F_{H_{\cal S}}(\rho_{\cal S})+\Delta \mathcal{E}_{{\cal SB}}+\beta^{-1}\left[D(\rho^{\prime}_{\cal B}||\rho_{{\cal B},\rm{th}})-D(\rho_{\cal B}||\rho_{{\cal B},\rm{th}})\right]\nonumber\\
   =&\Delta F+\Delta \mathcal{E}_{{\cal SB}}+\beta^{-1}\Delta D,
\end{align}
\end{widetext}
where $\Delta F=F_{H^{\prime}_{\cal S}}(\rho^{\prime}_{\cal S})-F_{H_{\cal S}}(\rho_{\cal S})$ is the change of free energy during the instant time interval, between two adjacent states \cite{adjacent}. Finally, we obtain an equality with incorporating information shown explicitly and naturally,
\begin{equation}\label{generalizedsecondlaw}
W-\Delta F-\Delta \mathcal{E}_{{\cal SB}}=\beta ^{-1}\left(\Delta I+ \Delta D\right),
\end{equation}
which extends the traditional second law of thermodynamics. We call it {\it the improved second law of thermodynamics}.

Obviously, the traditional form of the second law of thermodynamics $W-\Delta F\geq 0$ is restricted by three conditions: One is the product condition-the initial joint state of the system and surroundings should be a product state, so that there exists no initial correlations, which ensures $\Delta I\geq 0$. And then, the thermal equilibrium condition-the initial states of surroundings are in equilibrium thermodynamics, which  ensures $\Delta D\geq 0$. The end is the weak coupling condition-the couplings between the system and surrounding environment are weak, so that $\Delta \mathcal{E}_{{\cal SB}}\approx 0$, which ensures the least work $W$ is equal to the corresponding increase of free energy $\Delta F$.

Imposing the thermal equilibrium condition, it becomes
 \begin{align}\label{secondlawinter}
 W-\Delta F-\Delta \mathcal{E}_{{\cal SB}}&= \beta ^{-1}\big[\Delta I+D(\rho^{\prime}_{\cal B}||\rho_{{\cal B},\rm{th}})\big]\nonumber\\
&\geq \beta ^{-1}\Delta I,
 \end{align}
 which is the result discussed by Sagawa and Ueda using fluctuation theorem \cite{SagawaUeda2012}. The equality also is a generalized improvement of the result in \cite{Hasegawa2010} for a nonequilibrium initial state using the Jarzynski equality.

\subsection{Quantum information statement of the traditional principle of entropy increase}\label{QISL}
Energy conservation is essential and fundamental in the establishment of the three improved equality formulas. In essence, they reflect quantitative relationships between information and energy changes or transfers during the evolution \cite{furtherreading}. However, a confusing question arises: where is the direction of evolution? After all, the traditional second law concerns on directions of natural evolution processes. Without hesitation, the additional conditions which guarantee reductions to traditional forms, are suspects of the case, thus should to be reexamined.

One of the most noticeable features of the improved formulas Eqs.~\eqref{Improvedentropyincrease}, \eqref{improvedLandauerprinciple} and \eqref{generalizedsecondlaw} is the meaning on both sides of the equations, except the constant $k$ and the inverse temperature $\beta$, LH belongs to thermodynamics while RH belongs to information theory.
The least or common condition that guarantees reductions to traditional forms is the product condition, especially, in the formula of improved entropy increase, where the traditional principle is {\it identical} with the product condition. When adding more precisely and specially designed setups, other condition arises, e.g. the thermal equilibrium condition and the weak coupling condition.

A trustworthy interpretation is there has an identical corresponding expression of the traditional entropy increase of thermodynamics in quantum information theory, and then the direction of natural evolution process exists another meaning belongs to information theory. Thus, we have the following quantum information statement of the traditional principle of entropy increase.

{\it Proposition 1}\label{Proposition}: The evolution of an isolated quantum system, where no correlations exist, initially, between subsystems ${\cal S}_{\mu}, \mu=1,2,\ldots,N$, always towards to directions of the correlation information never be decreased, i.e. $I(\rho_{{\cal SB}}(t))\geq 0$.

It seems that Proposition 1 is not a result of new discovery, after all, the result $I(\rho_{{\cal SB}}(t))\geq 0$ can be proved by definitions of concepts of the correlation information for evolutions of an isolated quantum system, where energy conservation is satisfied. However, the significance is that we give the direction of natural evolution process a new meaning.

\section{Illustrations by physical examples}\label{sec.examples}
In this section, we illustrate the above obtained formulations through a physical example. A. Smirne et al. \cite{Smirne2010} used the Jaynes-Cummings model discussing initial correlations in open-system dynamics. The mode, especially when considering the initial correlations between the system and the bath reservoir, is suitable for what we want to illustrate in the above setups.

\begin{figure}[htbp]
\centering
\resizebox{0.44\textwidth}{!}{%
\includegraphics{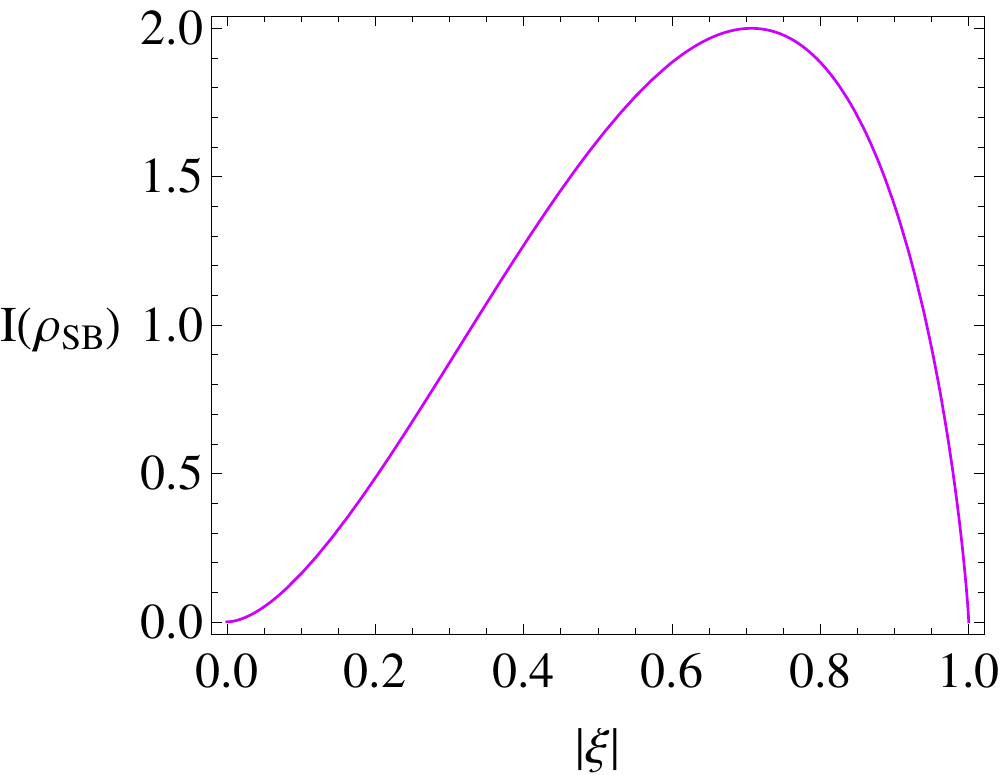}}
\caption{(Color online) The correlation contained in the initial state $\rho _{{\cal SB}}$, which characterized by the mutual information $I$, as a function of norm of the probability distribution coefficient $|\xi|$. The other parameters: $\omega_0=1$, $\omega=0.5$, $g=1$ and $n=7$.
 }\label{initialmutualinformation}
\end{figure}
\begin{figure*}[htbp]
\centering
\begin{minipage}[c]{0.44\textwidth}
\centering
\includegraphics[width=3in]{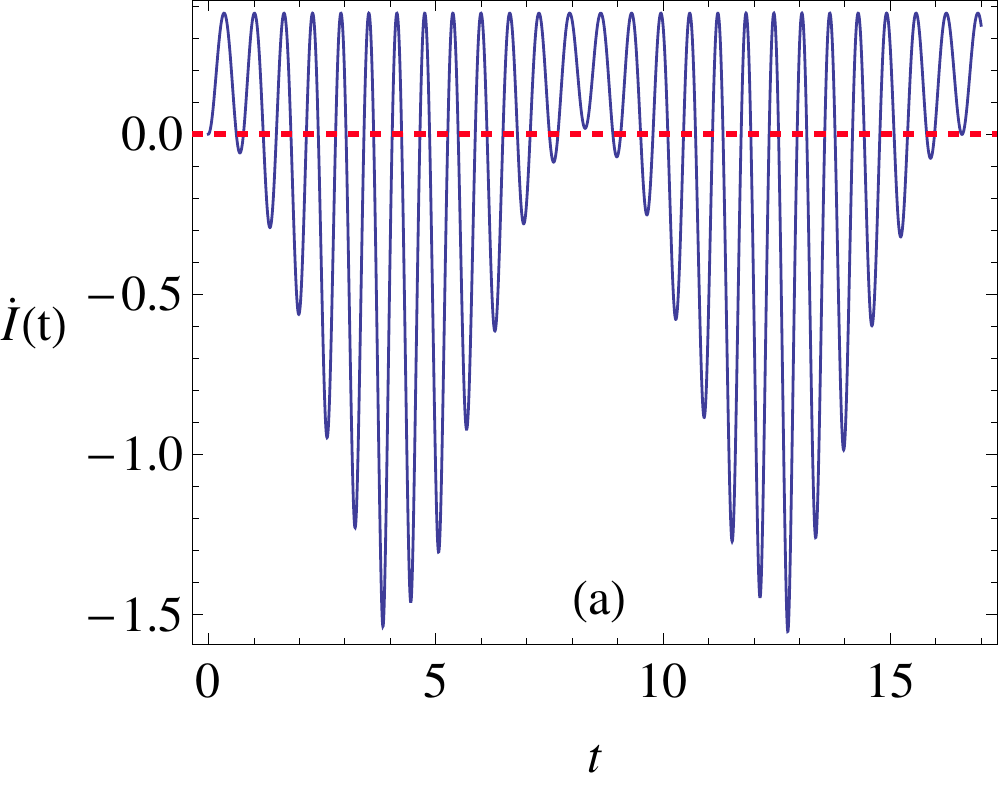}
\end{minipage}
\begin{minipage}[c]{0.44\textwidth}\centering
\includegraphics[width=3in]{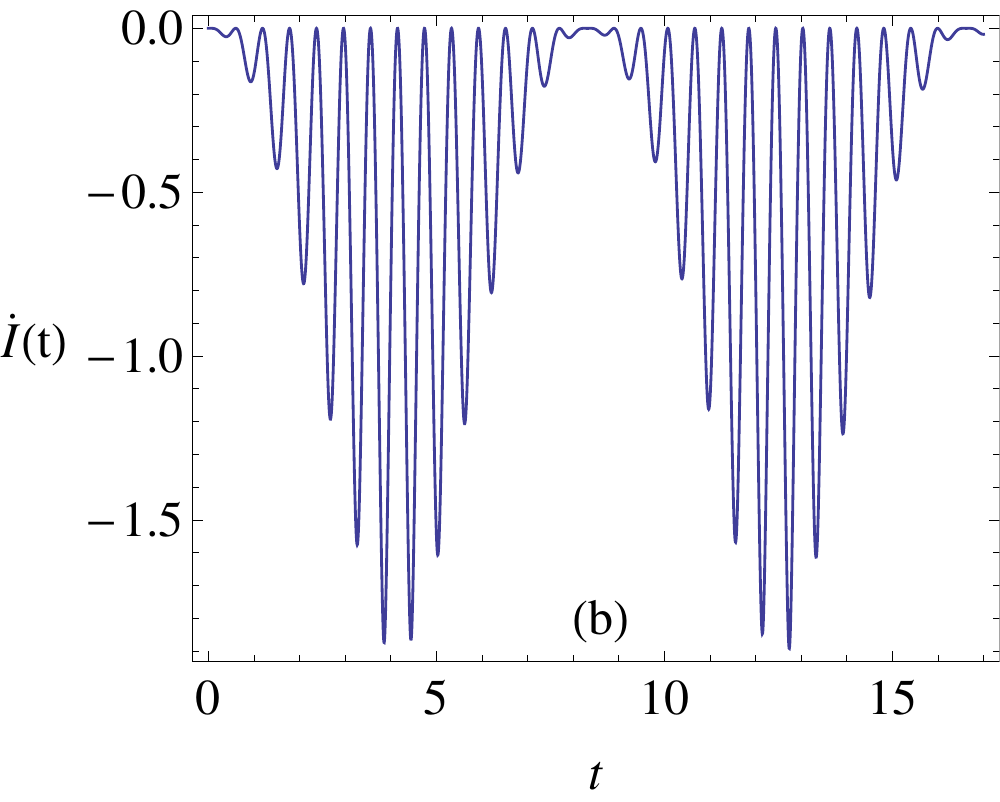}
\end{minipage}\\
\caption{(Color online) Plots of the change rate of the total entropy $\dot{I}(\rho_{{\cal SB}}(t))$. It illustrates that the change rate of the total entropy may be negative in some process of the dynamical evolutions. Parameters: $\omega_0=1$, $\omega=0.5$, $g=1$, $n=7$, with (a) $\xi= 0.5$ and (b) $\xi= 0.71$.
}\label{changeratetotalentropy}
\end{figure*}
\begin{figure}[htbp]
\centering
\resizebox{0.44\textwidth}{!}{%
\includegraphics{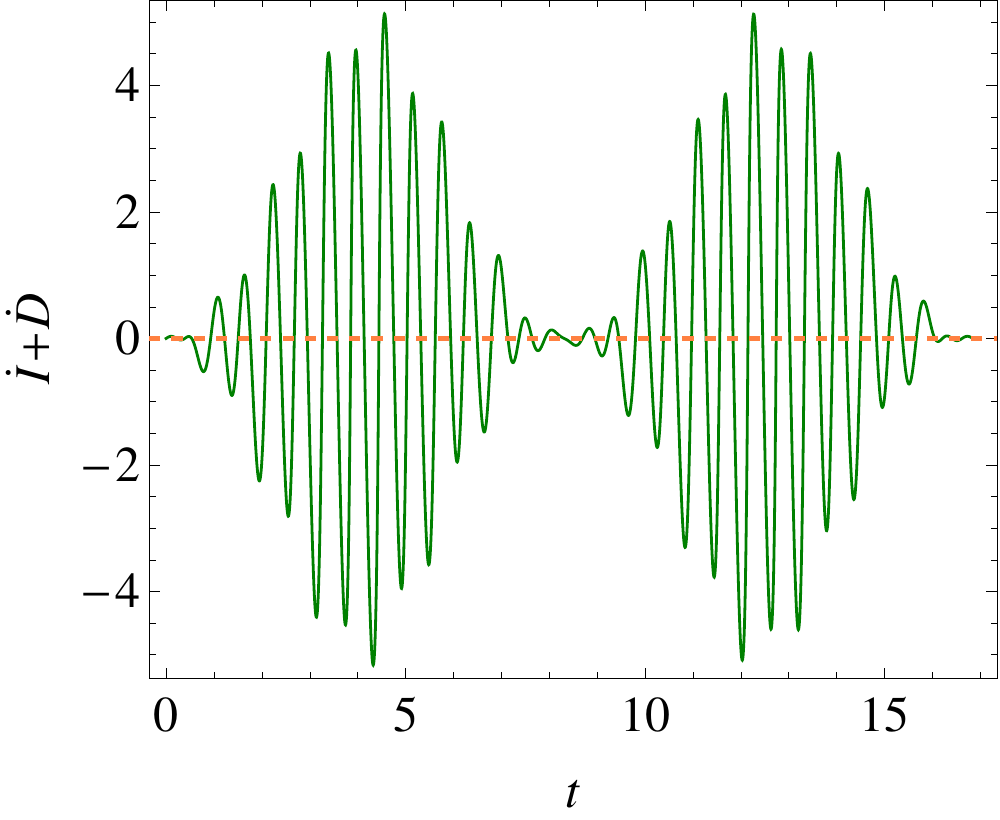}}
\caption{(Color online) Change rates of the mutual information plus the relative entropy of the bath reservoir as a function of time in dynamical evolutions, namely, $ \dot{I}+\dot{D}$. Parameters: $\xi= 0.71$, $\omega_0=1$, $\omega=0.5$, $g=1$, and $n=7$.
 }\label{changerate}
\end{figure}
\begin{figure*}[htbp]
\centering
\begin{minipage}[c]{0.44\textwidth}
\centering
\includegraphics[width=3in]{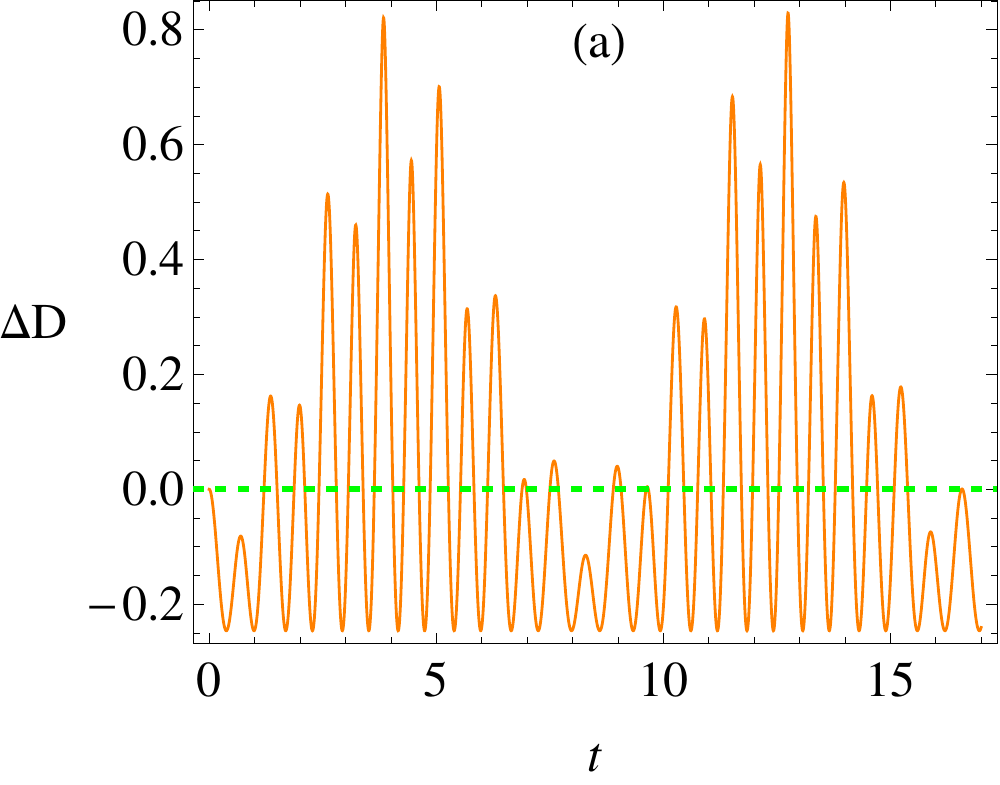}
\end{minipage}
\begin{minipage}[c]{0.44\textwidth}\centering
\includegraphics[width=3in]{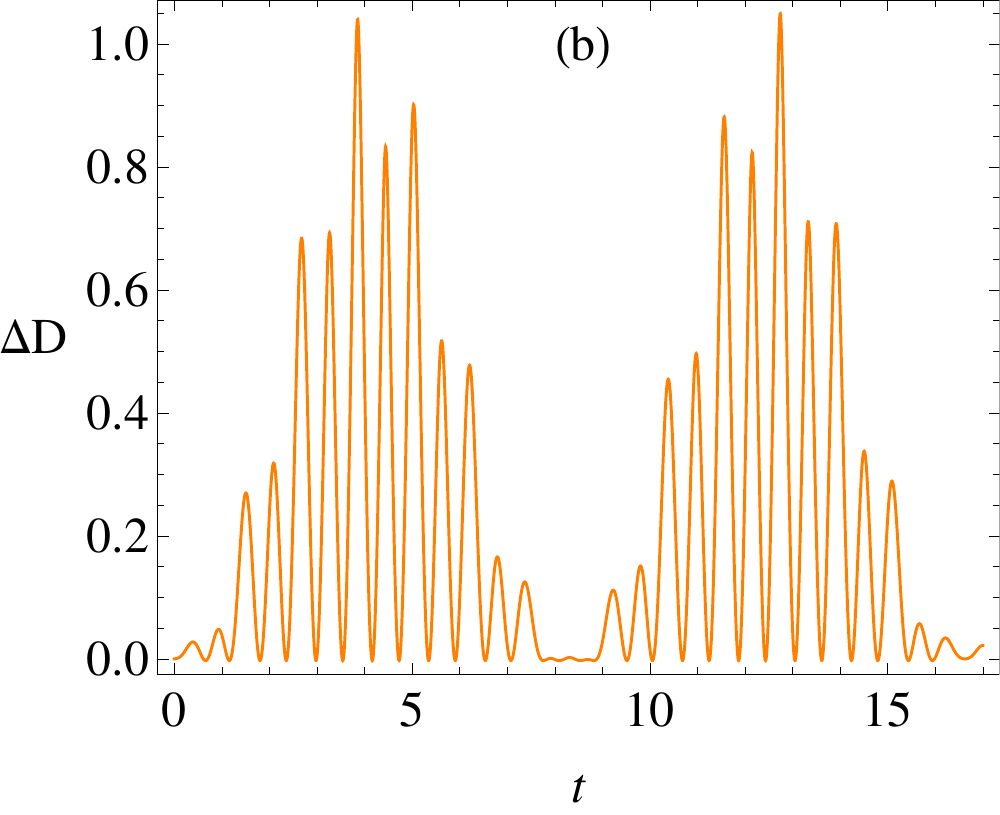}
\end{minipage}\\
\caption{(Color online) Plots of the increasing of the relative (information) entropy $\Delta D(t)$ of the bath reservoir with its initial thermal state as a function of time. Due to the non-equilibrium characteristics of the surroundings, it can be negative in evolution processes. Parameters: $\omega_0=1$, $\omega=0.5$, $g=1$, $n=7$, with (a) $\xi= 0.5$ and (b) $\xi= 0.71$.
}\label{changeraterelativeentropy}
\end{figure*}

Information are stored in a memory medium which implicated by a two-level system. The system in contact with a finite bath reservoir simulated by a single mode of the radiation field with total Hamiltonian
\begin{eqnarray} \label{JCHamiltonian}
  H &=& H_{\cal S} + H_{\cal B} + H_{{\cal SB}}\nonumber\\ &=& \omega_0 \sigma_+ \sigma_- + \omega b^{\dag} b + g
  \left( \sigma_+ \otimes b + \sigma_- \otimes b^{\dag} \right),
\end{eqnarray}
where $\sigma_+ = |1 \rangle \langle 0|$ and $\sigma_- = |0 \rangle \langle 1|$ are the raising and lowering operators of the two-level system, $b^{\dag}$ and $b$ are the creation and annihilation operators of the field mode. We consider an initially correlated pure state with the form
 \begin{equation} \label{initialstate}
 \rho_{{\cal SB}} (0) = | \psi \rangle \langle \psi | ,
\end{equation}
with $|\psi\rangle=\xi |0,n\rangle + \zeta |1,n-1\rangle$, $|\xi|^2+|\zeta|^2=1$. The evolution of state $\rho_{{\cal SB}} (t) $ is controlled by Eq.~\eqref{eqevolution} through an exact time-evolution operator for the total system in the interaction picture
\begin{equation} \label{timeevolutionoperator}
  U(t) = \left(\begin{array}{cc}
    c_{} ( \hat{n} + 1, t) & d_{} ( \hat{n} + 1, t) b\\
    - b^{\dag} d^{\dag} ( \hat{n} + 1, t) & c^{\dag} ( \hat{n}, t)
  \end{array}\right).
\end{equation}
Functions $c( \hat{n}, t)$ and $d(\hat{n}, t)$ of the number operator $\hat{n} = b^{\dag}b$ and detailed numerical calculations of the processing  are attached in Appendix \ref{appendix1}.

Using such special mode, initial correlations $I(\rho_{{\cal SB}})$, which characterized by the mutual information, are numerically illustrated in Fig.~\ref{initialmutualinformation} as a function of norm of the probability distribution coefficient $|\xi|$. For an isolated whole system, the total entropy Eq.~\eqref{Improvedentropyincrease} can be decreased in processes of the dynamical evolutions due to existence of initial correlations between the subsystems. We illustrated such abnormal characteristics in Fig.~\ref{changeratetotalentropy}. Initial correlations between the system and the bath reservoir also change the traditional Landauer's principle Eq.~\eqref{traditionalLandauer}. The flux equality $\beta  \dot{Q}\geq \dot{S}$ can not be maintained in process where change rate is negative, see Fig.~\ref{changerate}. This corresponds to the fact that the dynamical evolution draws the initial correlations (or informations), which not belonging to the reduced system and the bath reservoir individually, into the information processing. Fig.~\ref{changeraterelativeentropy} illustrates the increasing of the relative (information) entropy Eq.~\eqref{increasingrelen} of the bath reservoir with its initial thermal state as a function of time. Due to the non-equilibrium characteristics of the surroundings, it can be negative in evolution processes.

\section{Discussions}\label{conclusions}
In summary, in this paper, using the techniques of quantum statistical mechanics for thermodynamics, we recast relations between the fundament laws of thermodynamics and information theory generally. The initial correlations between the system and the surrounding environments are taken considered. From three different versions, all of them can be seen as  the second laws of thermodynamics, we obtained more general forms, which incorporate information explicitly and naturally. We exhibit the improved general forms as equalities, through which physical natures of information are clarified. Further, we relax restrictions on the initial state of bath reservoir, which can be far away from equilibrium. By virtue of law of energy conservation, all the forms be expressed as equality rather than inequality, which reflect quantitative relationships between information and energy changes or transfers during the evolution.

Giving the direction of natural evolution process a meaning totally belongs to information theory is a natural result when reducing the improved formulas to the tradition forms. An interesting theoretical stride by Vlatko Vedral \cite{Vedral2016}, recently, who treats the arrow of time as a manifestation of the increase of correlations, as quantified by the mutual information, between the system and the rest of the universe. Such a mode is an expanding and general discussion which we have give a foreshadowing in the discussion of improved principle of entropy increase.

Although, based on the setups, we have established a more generalized frame to connect the information theory and the fundamental laws of thermodynamics. However, many aspects are interesting and even challenging for subsequent works, such as, when measurements are taken into consideration, and also a time-independent Hamiltonian for the bath reservoir is a strong restriction on the evolutions. especially for small sized surroundings, where correlates with information register system are rapidly changing.

\begin{center}
\textbf{ACKNOWLEDGMENTS}
\end{center}
This work is supported by National Natural Science Foundation of China (Nos.11547050, 91536108), Natural Science Foundation of Shaanxi Province (No.2016JM1027) and CAS (XDB01010000, XDB21030300), and partially supported by  MOST (2016YFA0302104, 2016YFA0300600). We thank useful discussions with Vlatko Vedral and Mile Gu.

\appendix
\section{}\label{appendix1}

The functions of the number operator $\hat{n}=b^{\dag}b$ have the form
\begin{eqnarray}\label{eq:cd}
  c ( \hat{n}, t) & = & e^{i \Delta t / 2}  \left[ \cos \left( \Omega (
  \hat{n})_{} \frac{t}{2} \right) - i \frac{\Delta}{\Omega ( \hat{n})} \sin
  \left( \Omega ( \hat{n}) \frac{t}{2} \right) \right] ,\nonumber\\
  d ( \hat{n}, t) & = & - ie^{i \Delta t / 2} \frac{2 g}{\Omega ( \hat{n})}
  \sin \left( \Omega ( \hat{n}) \frac{t}{2} \right),
\end{eqnarray}
with
\begin{equation} \label{eq:omn}
  \Omega (\hat{n}) = \sqrt{\Delta^2 + 4 g^2 \hat{n}},
\end{equation}
where $\Delta=\omega_0-\omega$ denotes the detuning between the system's transition frequency $\omega_0$ and the frequency
$\omega$ of the field mode.

One can expand the initial state Eq.~\eqref{initialstate} using the basis vector $|\mu\rangle\otimes|m\rangle\equiv|\mu,m\rangle$, with $\mu=0,1$ labels the states of the two-mode system and $m=0,1,2,\ldots$ the number states of the field mode,
\begin{equation}\label{expandinitialstate}
    \rho_{{\cal SB}}(0)=\sum_{\mu, \nu, m, l}\rho^{m,l}_{\mu,\nu}|\mu,m\rangle \langle \nu,l|.
\end{equation}
The non-trivial  coefficients are
\begin{equation}\label{coefficients}
\begin{aligned}
\rho^{m,l}_{00}&=|\xi|^2\delta_{mn}\delta_{ln},\\
\rho^{m,l}_{01}&=\xi\zeta^*\delta_{mn}\delta_{l,n-1} \\
\rho^{m,l}_{10}&=\xi^*\zeta\delta_{m,n-1}\delta_{l,n},\\
\rho^{m,l}_{11}&=|\zeta|^2\delta_{m,n-1}\delta_{l,n-1}.
\end{aligned}
\end{equation}
Using the evolution equation Eq.~\eqref{eqevolution}, after some calculations one can obtained the reduced states
\begin{align}
    \rho_{\cal S}(t)&=\left(
                \begin{array}{cc}
                  a_1+b_1 &0 \\
                  0 & a_2+b_2 \\
                \end{array}
              \right),\label{reducedstateS01}\\
    \rho_{\cal B}(t)&=\left(
                  \begin{array}{cccc}
                    a_1 & 0 & a_3 & 0 \\
                    0 & a_2 & 0 &b_3 \\
                    a^*_3 & 0 & b_1 & 0 \\
                    0 & b^*_3 & 0 & b_2 \\
                  \end{array}
                \right)\label{reducedstateS02},
\end{align}
for system and bath reservoir respectively. The elements are
\begin{equation}\label{reducedstateS03}
\begin{aligned}
a_1&=(n-1)|\zeta|^2|d_{n-1}(t)|^2,\\
a_2&=|\zeta|^2|c_{n-1}(t)|^2,\\
a_3&=\sqrt{n-1} \xi^*\zeta c^*_{n+1}(t) d_{n-1}(t),\\
b_1&=|\xi|^2|c_{n+1}(t)|^2,\\
b_2&=(n+1) |\xi|^2|d_{n+1}(t)|^2,\\
b_3&=-\sqrt{n+1} \xi^*\zeta c^*_{n-1}(t) d_{n+1}(t),
\end{aligned}
\end{equation}
where $c_n(t )$ and $d_n(t )$ denote the eigenvalues of $c(\hat n, t)$ and $d(\hat n, t)$, corresponding to the eigenstate $|n\rangle$, respectively. Here, one should note that base vectors of states Eqs.~\eqref{reducedstateS01} and \eqref{reducedstateS02} are \{$|\mu\rangle$, $\mu=0, 1$\} for the two-mode system and \{$|m\rangle$ ,$m=n-2, n-1, n, n+1$\} for the bath reservoir, respectively.

Now, by virtue of detailed expressions of states Eqs.~\eqref{reducedstateS01} and \eqref{reducedstateS02}, the change of the total entropy $\Delta I$ can be given out using Eq.~\eqref{Improvedentropyincrease} and the definition of von Neumann entropy  Eq.~\eqref{vonNeumannentropy}. Similarly, one can obtain the rate of changes $\dot{I}\equiv \rm{d}I/\rm{d} t$.

Conveniently, we rewrite the state of bath reservoir Eq.~\eqref{reducedstateS02} as
\begin{equation}\label{reducedstateS021}
\begin{aligned}
\rho_{\cal B}(t)=~~& a_1 |n-2\rangle \langle n-2|+a_3 |n-2\rangle \langle n|\\
+&a_2 |n-1\rangle \langle n-1|+b_3 |n-1\rangle \langle n+1|\\
+&a^*_3 |n\rangle \langle n-2|+b_1 |n\rangle \langle n|\\
+&b^*_3 |n+1\rangle \langle n-1|+b_2 |n+1\rangle \langle n+1|.
\end{aligned}
\end{equation}
Using the above form, the energy of the bath reservoir is
\begin{equation}\label{energybathreservoir}
E(t)=\tr{H_{\cal B} \rho_{\cal B}(t)}=\tr{\omega b^{\dag}b\rho_{\cal B}(t)}.
\end{equation}
Further, we have the inverse temperature
\begin{equation}\label{inversetemperature}
\beta(t)=\frac{1}{\omega} \ln{\left(1+\frac{\omega}{E(t)}\right)}
\end{equation}
for the thermal state  $\rho_{{\cal B},\rm{th}}=e^{-\beta H_{\cal B}}/\tr{e^{-\beta H_{\cal B}}}$, which corresponds to the state of the bath reservoir $\rho_{\cal B}(t)$ and they have the same energy. In order to facilitate the calculation of $D(\rho^{\prime}_{\cal B}||\rho_{{\cal B},\rm{th}})$, similar with Eq.~\eqref{reducedstateS021} we re-express the thermal state in a matrix form
\begin{equation}\label{thermalstatematrix}
\rho_{{\cal B},\rm{th}}(t)=\frac{1}{Z}\sum_{m}\rm{e}^{-\beta(t) m \omega}|m\rangle\langle m|,
\end{equation}
with the partition function $Z=1/\left [1-\rm{e}^{-\beta(t) m \omega}\right]$.

Straightforwardly,  using Eqs.~\eqref{reducedstateS021} and \eqref{thermalstatematrix} one can obtain the final result of the relative entropy $D(\rho^{\prime}_{\cal B}||\rho_{{\cal B},\rm{th}})$ and time rate of changes $\dot{D}\equiv \rm{d}D/\rm{d} t$. In the main text, we plot the rate of changes of $\dot{I}+\dot{D}$ as a function of time.

\newcommand{\PR}{Phys. Rev. }
\newcommand{\PRL}{Phys. Rev. Lett. }
\newcommand{\PRA}{Phys. Rev. A }
\newcommand{\PRE}{Phys. Rev. E }
\newcommand{\PRX}{Phys. Rev. X }
\newcommand{\JPA}{J. Phys. A }
\newcommand{\JPB}{J. Phys. B }
\newcommand{\PLA}{Phys. Lett. A }
\newcommand{\NP}{Nat. Phys. }
\newcommand{\NC}{Nat. Commun. }
\newcommand{\RMP}{Rev. Mod. Phys. }



\begin{thebibliography}{99}
\bibitem{adjacent}  Here, the word ``adjacent" comes form understanding of the free energy $F_{H_{\cal S}{^\prime}}$, where the meaning of inverse temperature $\beta:=\beta(t_0)$ is corresponding to the initial moment $t_0$, but not to $t$. This is reasonable and coincident with the definition, since we are considering a process over a limited time interval $\Delta t=t-t_0$. That is, we have ignored the second order small quantities.
\end{thebibliography}
\end{document}